\documentclass[useAMS,usenatbib]{mnras}
\usepackage{graphicx}
\usepackage{epstopdf}
\usepackage{amsmath,esint,bm}
\usepackage{amssymb}

\usepackage{etoolbox}
\makeatletter
\patchcmd\@combinedblfloats{\box\@outputbox}{\unvbox\@outputbox}{}{%
   \errmessage{\noexpand\@combinedblfloats could not be patched}%
}%
 \makeatother

\title[Gauss's Law and Poisson's Equation for Varying G]{Gauss's Law and the Source for Poisson's Equation in Modified Gravity with Varying G}
\author[Christodoulou \& Kazanas]{Dimitris M. Christodoulou$^{1,2}$  and Demosthenes Kazanas$^{3}$
\\
$^{1}$Lowell Center for Space Science and Technology, University of Massachusetts Lowell, Lowell, MA, 01854, USA.\\
$^{2}$Dept. of Mathematical Sciences, Univ. of Massachusetts Lowell,
Lowell, MA, 01854, USA. E-mail: dimitris\_christodoulou@uml.edu\\
$^{3}$NASA/GSFC, Laboratory for High-Energy Astrophysics, Code 663, Greenbelt, MD 20771, USA. E-mail: demos.kazanas@nasa.gov\\
}

\begin{document}

\def\gsim{\mathrel{\raise.5ex\hbox{$>$}\mkern-14mu
                \lower0.6ex\hbox{$\sim$}}}

\def\lsim{\mathrel{\raise.3ex\hbox{$<$}\mkern-14mu
               \lower0.6ex\hbox{$\sim$}}}

%\date{Accepted ??? . Received 2017 March ??; in original form 2017 March ??}
\pagerange{\pageref{firstpage}--\pageref{lastpage}} \pubyear{2018}

\maketitle

\label{firstpage}

\begin{abstract}
We have recently shown that the baryonic Tully-Fisher and Faber-Jackson relations imply that the gravitational ``constant'' $G$ in the force law varies with acceleration $a$ as $G\propto 1/a$ and vice versa. These results prompt us to reconsider every facet of Newtonian dynamics. Here we show that the integral form of Gauss's law in spherical symmetry remains valid in $G(a)$ gravity, but the differential form depends on the precise distribution of $G(a)M(r)$, where $r$ is the distance from the origin and $M(r)$ is the mass distribution. We derive the differential form of Gauss's law in spherical symmetry, thus the source for Poisson's equation as well. Modified Newtonian dynamics (MOND) and weak-field Weyl gravity are asymptotic limits of $G(a)$ gravity at low and high accelerations, respectively. In these limits, we derive telling approximations to the source in spherical symmetry. It turns out that the source has a strong dependence on surface density $M/r^2$ everywhere in $a$-space except in the deep Newton-Weyl regime of very high accelerations.
\end{abstract}




\section{Introduction}\label{intro}

In \cite{chr18a,chr18b}, we showed that in the regime in which the observed baryonic Tully-Fisher \citep{tul77,mcg00,mcg12} and Faber-Jackson  \citep{fab76,san09,den15} relations are valid, the gravitational ``constant'' $G$ should vary with acceleration $a$ in the force law. Such a varying $G(a)$ function can naturally account for the non-Newtonian force postulated in Modified Newtonian Dynamics (MOND) \citep{mil83a,mil83b,mil83c,mil15a,mil15b,mil16}, as well as for additional terms that appear only in weak-field Weyl gravity \citep{man89,man94}.

Weak Weyl gravity (WG) and MOND are asymptotic limits in varying $G$ gravity and, as such, they do not have the benefit of relying on a fundamental principle such as the precise variation of $G(a)$ at all acceleration regimes. For this reason, the asymptotic cases face difficulties in both deciphering the origin of their equations and identifying the embedded universal constants. In particular, the gravitational constant $G_0$ does not appear in the metric of WG \citep{man89,man94} and a mysterious scale-invariant constant $G_0a_0$ appears in MOND, where there is also no unique Poisson equation and speculations abound \citep{fam12,mil15c}. Here $G_0$ is the Newtonian constant value of $G$ and $a_0$ is a transitional constant acceleration below which gravity is decidedly non-Newtonian.

The empirical freedoms of assigning constants by inspection and writing down various Poisson equations do not permit the above asymptotic cases to mature and develop into complete theories of gravity, whereas $G(a)$ gravity is much more constrained and enjoys no such freedoms. In this work, we study the details of the spherically symmetric $G(a)$ gravity. We show that the modified force law $a=G(a) M(r)/r^2$, where $M$ is mass and $r$ is distance from the origin, implies that the integral form of Gauss's law remains valid, but the differential form of the law needs to be modified. Specifically, the divergence of the gravitational field and the Laplacian of Poisson's equation require a brand new source which does not depend exclusively on the gradient $dM(r)/dr$ or, equivalently, on the volume density $\rho(r)$ of the mass distribution. By and large, the source depends on the surface density $\sigma\equiv M(r)/r^2$ and this fundamental dependence does not go away even in the simplest case of a mere point-mass $M=$ constant. The important role that $\sigma$ plays in the dynamics of astrophysical objects has been recognized previously, both empirically in MOND (where the constant $a_0/G_0$ has dimensions of surface density) and observationally by looking at all scales in the universe \citep[e.g.,][]{lar81,kaz95,mil16}. But no explanation could be offered until now. As we show in this work, the dominant contribution of the surface density that cannot be removed from the equations is independent of the mass distribution and a manifestation of $G(a)$ gravity with $G(a) = G_0 + G_0 a_0/a$. This is a ``bare minimum'' formula adopted because it behaves correctly in the two asymprotic cases $a\gg a_0$ (WG) and $a\ll a_0$ (MOND) and, unlike MOND's interpolating functions, it does not introduce any additional spurious forces in the regime of intermediate accelerations \citep{chr18a,chr18b}.

In \S~\ref{acc}, we discuss the details of Gauss's law and we derive the source term for Poisson's equation in the general case (\S~\ref{Q}) as well as in the special cases with $M =$ constant and $\rho =$ constant (\S~\ref{cases}), and $\sigma =$ constant (\S~\ref{csd}). In \S~\ref{derivation}, we derive the asymptotic limits of the new source term for $a\gg a_0$ (Newton-Weyl) and $a\ll a_0$ (MOND). In \S~\ref{conc}, we summarize and discuss our results.

\section{The source in Gauss's Law with Varying ${\mathbf G}$ in Spherical Symmetry}\label{acc}

\subsection{The Functions ${{\mathbf a(s)}}$ and ${{\mathbf G(s)}}$}\label{gofs}

For all accelerations irrespective of magnitude $a$, the function $G(a)$ is constrained by the asymptotic cases to take the specific form
\begin{equation}
G(a) = G_0  + \frac{G_0a_0}{a}\, ,
\label{xG1}
\end{equation}
where $G_0 = 6.674\times 10^{-11}$ m$^3$ kg$^{-1}$ s$^{-2}$
is the Newtonian gravitational constant and 
$$G_0a_0 = 8.0\times 10^{-21}~{\rm m}^4\,{\rm kg}^{-1}\,{\rm s}^{-4}\, ,$$
is a new characteristic constant with dimensions $[v]^4/[M]$ that naturally appears in the deep MOND limit as well \citep{mil15c}.

Force balance has produced the following expression for the acceleration $a$ \citep{mcg16,lel17,chr18a}: 
\begin{equation}
a = \frac{G(a)M(r)}{r^2} = \frac{a_N}{2}\left(1 + \sqrt{1 + \frac{4a_0}{a_N}}\right) \, ,
\label{f2}
\end{equation}
where the Newtonian acceleration $a_N$ is defined by
\begin{equation}
a_N \equiv \frac{G_0M(r)}{r^2}\, .
\label{f3}
\end{equation}
A casual look at these equations reveals that $a$ is a function of radial distance $r$. This casual observation has hindered progress in the field. A closer look shows that $a$ and $a_N$ are both functions of the surface density $\sigma\equiv M(r)/r^2$, a neglected property that is not clearly visible in MOND either, despite the fact that MOND detects a new characteristic surface density of the form $\sigma_0 = a_0/G_0$. By defining the dimensionless surface-density parameter $s\equiv \sigma /\sigma_0$, we rewrite eq.~(\ref{f2}) in the form 
\begin{equation}
a(s) = \frac{a_0}{2}\left( s + \sqrt{s^2 + 4s} \right) \, ,
\label{f4}
\end{equation}
and eq.~(\ref{xG1}) in the form
\begin{equation}
G(s) = G_0\left( 1 + \frac{2}{s + \sqrt{s^2 + 4s}} \right) \, .
\label{f5}
\end{equation}
It is now evident that the surface density $s\propto M(r)/r^2$, not the mass $M(r)$ or the radius $r$ individually, is in complete control of the dynamics in spherical symmetry. Not only does $s$ specify how the acceleration should vary, but it also controls the variation of $G(s)$ itself.\footnote{Furthermore, in the critical case $s=1$, both $a/a_0$ and $G/G_0$ are exactly equal to the golden ratio $\varphi = (1+\sqrt{5})/2\approx 1.618$. It is interesting that this ratio that does appear in classical Newtonian dynamics \citep{chr17a,chr17b} is not absent from $G(a)$ gravity either.} This is a brand new revelation and it allows us to proceed and study additional fundamental principles in varying $G$ gravity such as Gauss's law and Poisson's equation.

In what follows, we will also make use of the auxiliary equations
\begin{equation}
\frac{ds}{dr} = \frac{1}{\sigma_0}\left(\frac{1}{r^2}\frac{dM}{dr}\right) - \frac{2s}{r}
\, ,
\label{s1}
\end{equation}
where $\sigma_0\equiv a_0/G_0 = 1.8$~kg~m$^{-2}$ = 860~$M_\odot$~pc$^{-2}$ \citep[in contemporary parlance, $a_0/(2\pi G_0) = 137$~$M_\odot$~pc$^{-2}$; see, e.g.,][]{don09},
and
\begin{equation}
\frac{dG}{ds} = \frac{-G_0}{s\sqrt{s^2 + 4s}}\, ,
\label{s2}
\end{equation}
that are derived from the above $s$-dependent equations.

\subsection{Integral Form of Gauss's Law}\label{if}

The integral form of Gauss's law 
\begin{equation}
\oiint_S{ {\boldsymbol a}\,{\boldsymbol\cdot}\,d{\bf S} } = 4 \pi G M \, ,
\label{g1}
\end{equation}
is valid in $G(a)$ gravity for a spherical Gaussian surface $S$ of radius $r$. Eq.~(\ref{g1}) is equivalent to the force law shown in eq.~(\ref{f2}), that is $a=G(a)M(r)/r^2$.

\subsection{Differential Form of Gauss's Law}\label{df}

The differential form of Gauss's law is derived from eq.~(\ref{g1}) and the divergence theorem; it takes the form
\begin{equation}
\iiint_V{(\boldsymbol{\nabla\cdot a}})\, dV = 4 \pi G M \, ,
\label{g2}
\end{equation}
where $V$ is a spherical volume of radius $r$. Since $G(a)$ varies implicitly with radius $r$, the right-hand side cannot be trivially written as a volume integral. We rewrite eq.~(\ref{g2}) as
\begin{equation}
\iiint_V{(\boldsymbol{\nabla\cdot a}})\, dV = 4 \pi G M \equiv \iiint_V{Q(r)\, dV} \, ,
\label{g4}
\end{equation}
where the source function $Q(r)$ is to be determined from the equation
\begin{equation}
Q(r) = \frac{d}{dV}\left(4\pi G M\right) = \frac{1}{r^2}\frac{d}{dr}\Big[G(a)M(r)\Big] \, .
\label{g5}
\end{equation}
Then, the differential form~(\ref{g4}) of Gauss's law can be written as
\begin{equation}
\boldsymbol{\nabla\cdot a} = Q \, ,
\label{g6}
\end{equation}
instead of the familiar form $\boldsymbol{\nabla\cdot a} = 4 \pi G_0\rho$. We note that, in spherical symmetry, eq.~(\ref{g6}) is equivalent to eq.~(\ref{g5}).

The function $Q$ also serves as the source in Poisson's equation: using the definition $a\equiv -\boldsymbol{\nabla}\Phi$ for the gradient of the gravitational potential $\Phi$, eq.~(\ref{g6}) takes the form 
\begin{equation}
\nabla^2\Phi = -Q \, .
\label{p1}
\end{equation}
This form of Poisson's equation, with $Q$ given by eq.~(\ref{g5}), is unique and binding in $G(a)$ gravity. No modifications are allowed such as those that were proposed in the context of MOND (various versions of MOND are free to modify the inertial mass in the Newtonian force law, or the Poisson equation itself, or both). On the other hand, $G(a)$ gravity assumes that $G$ varies in space and then, there is no more freedom in toying with the intertial mass or choosing conveniently the forms of forces and potentials. Thus, $G(a)$ gravity is a more restrained modified gravity case than MOND and it just happens that the MOND force law appears as an asymptotic limit for very low accelerations and surface densities. In $G(a)$ gravity, we simply cannot allow for such freedoms in the force law or in Poisson's equation; eq.~(\ref{g5}) provides a strong constraint in the spherical case.

\subsection{The Source Function ${{\mathbf Q(r)}}$}\label{Q}

The source function $Q(r)$ is determined from eq.~(\ref{g5}). In special cases in which the force law is known, eq.~(\ref{g6}) can also be used to determine $Q$. This serves as an independent check of the derivations that follow and we have confirmed the agreement between results for the special cases with $M =$ constant, $\rho =$ constant, and $\sigma =$ constant.

Expanding the right-hand side of eq.~(\ref{g5}), we find that
\begin{eqnarray*}
Q(r) & = & \sigma \frac{dG}{dr} + \frac{G}{r^2} \frac{dM}{dr} \\
       & = & \sigma_0 s \frac{ds}{dr} \frac{dG}{ds} + \frac{G}{r^2} \frac{dM}{dr}\, . \\
\end{eqnarray*}
Substituting eqs.~(\ref{s1}) and~(\ref{s2}) in the last equation above and grouping together all terms that depend on $dM/dr$, we find that
\begin{equation}
Q(r) = \frac{2a_0}{r} \frac{s}{\sqrt{s^2+4s}} + \frac{G_0}{2r^2} \frac{dM}{dr}\left(1 + \frac{s+2}{\sqrt{s^2+4s}} \right) \, .
\label{q2}
\end{equation}
The Newtonian dependence of the source on $dM/dr\propto\rho(r)$ is still present, but the gradient of the mass distribution is now coupled to the surface density $s$. In particular, the term in parentheses is proportional to the gradient $da/ds$, a surprising new dependence that does not occur in Newtonian theory.
Furthermore, the first term on the right-hand side is independent of mass and it does not drop out in any special case. This Weyl-like term is contributed by $dG/dr$ itself, irrespective of the underlying distribution of mass or any other factors. Finally, the constants $a_0$ and $G_0$ appear to be decoupled, but this is temporary and due to the units used in this equation. (Written in this convenient form, $Q$ has dimension of $[T]^{-2}$.)

In eq.~(\ref{q2}), we see a dependence of the source term on the mass gradient $dM/dr$. This type of dependence is not new. Extensions of Horndeski
scalar-tensor relativistic theory \citep[e.g.,][]{kob15} also predict a dependence of Poisson's equation on derivatives of the mass distribution. This dependence can actually be used to differentiate between various competing theories of modified gravity. This conclusion is unexpected, it shows a way to examine competing modifications of gravity in the weak field limit and choose the one that fits spherical systems best.

Eq.~(\ref{q2}) can be written in an alternative form with consistent units in both terms on the right-hand side as follows: We replace mass from the definition $M=\sigma r^2$; its gradient is
\begin{equation}
\frac{dM}{dr} = \sigma_0 \frac{d}{dr}\Big(r^2 s\Big) \, ,
\label{q3}
\end{equation}
and $Q(r)$ then becomes
\begin{equation}
Q(r) = \frac{2a_0}{r} \frac{s}{\sqrt{s^2+4s}} + \frac{a_0}{2r^2} \frac{d}{dr}\Big(r^2 s\Big)\left(1 + \frac{s+2}{\sqrt{s^2+4s}} \right) \, .
\label{q4}
\end{equation}
We note that we cannot scale $r$ in this equation by some characteristic length because the length scale depends on the distribution of mass that has to be specified first. This is done in the special cases that follow and it leads to different length scales depending on the adopted $M(r)$.

\subsection{Special Cases}\label{cases}

(a) $Point$-$mass~M = constant$. In the case of a point-mass, the surface density reduces to $s = 1/x^2$, where $x\equiv r/r_M$ and the characteristic length scale is $r_M=\sqrt{G_0M/a_0}$. Since $dM/dr=0$ for $r\neq 0$, the mass $M$ only sets the length scale and it does not affect $Q$ otherwise.
With these equations, eq.~(\ref{q2}) reduces to
\begin{equation}
Q(x) = \frac{a_0}{r_M}\left( \frac{2}{x\sqrt{1+4x^2}}\right) \, .
\label{q5}
\end{equation}
The same result can be obtained directly from eq.~(\ref{g6}) by substituting eqs.~(\ref{f4}) and~(\ref{s1}) with $dM/dr=0$. \\

\noindent
(b) $Density~\rho = constant$. For an extended spherical mass distribution, its gradient in eq.~(\ref{q4}) reduces to
\begin{equation}
\frac{a_0}{2r^2} \frac{d}{dr}\Big(r^2 s\Big) = 2\pi G_0 \rho(r)\, ,
\label{q6}
\end{equation}
and $\rho(r)$ is responsible for setting the length scale $L$ of the distribution. For a uniform distribution $\rho =\rho_0 =$ constant, and then $L$ is found to be
\begin{equation}
L = \frac{3 a_0}{4\pi G_0\rho_0}\, .
\label{q7}
\end{equation}
Then, $s=r/L$ and eq.~(\ref{q4}) reduces to
\begin{equation}
Q(s) = 2\pi G_0 \rho_0\left(1 + \frac{s+10/3}{\sqrt{s^2+4s}}\right)\, .
\label{q8}
\end{equation}
The same result can be obtained directly by substituting eq.~(\ref{f4}) and $r/L=s$ into eq.~(\ref{g6}). In the deep Newtonian limit, $s\to\infty$, the fraction in the parentheses approaches 1 and the familiar Poisson's equation, $\nabla^2\Phi = -4\pi G_0\rho_0$, is obtained from eqs~(\ref{q8}) and~(\ref{p1}).

\subsection{The Case of Constant Surface Density}\label{csd}

In the case of constant surface density ($\sigma =$ const., $s =$ const.), Poisson's equation can be solved exactly for all accelerations, both in the pure Newtonian case and in $G(a)$ gravity. In both cases, the acceleration is constant and the gravitational potential is precisely the same. This means that a spherically symmetric mass distribution with $M/r^2 =$ constant is {\it oblivious} to the variation of $G(a)$ described by eq.~(\ref{xG1}).
\\

\noindent
(a) $Pure~Newtonian~Case~(G\equiv G_0)$: For $\sigma = M(r)/r^2 =$ constant, the Newtonian acceleration takes the form
\begin{equation}
a_N =G_0 M(r)/r^2 = G_0\, \sigma = a_0\, s = {\rm constant}\, ,
\label{A1}
\end{equation}
where $s\equiv \sigma/\sigma_0$ and $\sigma_0 \equiv a_0/G_0$. Next we write $a_N \equiv d\Phi/dr$, where $\Phi(r)$ is the gravitational potential defined without the negative sign so that $a>0$, and then we find by integration that
\begin{equation}
\Phi(r) = a_0\, s\, r = a_N\, r\, ,
\label{A2}
\end{equation}
thus the potential $\Phi(r)$ is linear in $r$ and the constant $a_N$ serves as the slope.
\\

\noindent
(b) $G(a)~Gravity$: For $\sigma = M(r)/r^2 =$ constant, the $G(a)$ acceleration takes the form
\begin{equation}
a =G(a) M(r)/r^2 = G(a)\, \sigma\, ,
\label{A3}
\end{equation}
which implies that $a =$ constant in this case as well. Then, combining eq.~(\ref{p1}) (without the negative sign so that $a~>~0$) with eq.~(\ref{q2}), we obtain Poisson's equation in spherical symmetry, viz.
\begin{equation}
\frac{1}{r^2}\frac{d}{dr}r^2\frac{d\Phi}{dr} = \frac{2a_0}{r}f(s) + \frac{G_0}{2r^2}\frac{dM}{dr}g(s)\, ,
\label{A4}
\end{equation}
where the functions $f(s)$ and $g(s)$ are defined in \S~\ref{derivation} below. Integrating twice and using the boundary conditions that (i) $\Phi(0)=0$ and (ii) $d\Phi/dr$ remains finite at the origin, we find that
\begin{equation}
\Phi(r) = a_0\left(f(s) + \frac{s}{2}g(s) \right)r \equiv a\, r\, ,
\label{A5}
\end{equation}
where $a$ is given by eq.~(\ref{f4}) above and $a =$ constant since $s =$ constant. In this case as well, the potential $\Phi(r)$ is linear in $r$ and the constant $a$ serves as the slope.

\section{Asymptotic Forms}\label{derivation}

\subsection{The $s\to 0$ and $s\to\infty$ limits}

Eqs.~(\ref{q2}) and~(\ref{q4}) contain two functions that depend on the dimensionless surface density $s$, namely
\begin{equation}
f(s) = \frac{s}{\sqrt{s^2+4s}} \, ,
\label{s01}
\end{equation}
and
\begin{equation}
g(s) = 1 + \frac{s+2}{\sqrt{s^2+4s}} \, .
\label{s02}
\end{equation}
Their series expansions are as follows: In the deep MOND limit, $s\to 0$,
\begin{equation}
f(s) \approx \frac{\sqrt{s}}{2}\left(1 - \frac{s}{8} + \frac{3s^2}{128} - \frac{5s^3}{1024}\right) \, ,
\label{s03}
\end{equation}
and
\begin{equation}
g(s) \approx \frac{1}{\sqrt{s}}\left(1 + \sqrt{s} + \frac{3s}{8} - \frac{5s^2}{128} + \frac{7s^3}{1024}\right) \, .
\label{s04}
\end{equation}
Thus, $g\gg f$ in the MOND regime and the Weyl-like term $f(s)$ is negligible.
In the Newtonian limit, $s\to\infty$,
\begin{equation}
f(s) \approx 1 - \frac{2}{s} + \frac{6}{s^2} - \frac{20}{s^3} + \frac{70}{s^4} \, ,
\label{s05}
\end{equation}
and
\begin{equation}
g(s) \approx 2\left(1 + \frac{1}{s^2} - \frac{4}{s^3} + \frac{15}{s^4}\right) \, .
\label{s06}
\end{equation}
Thus, $f\approx 1$ and $g\approx 2$ in the deep Newtonian limit.

\subsection{Zeroth-order Approximations}

The zeroth-order approximations for $Q$ can be obtained from eq.~(\ref{q2}) and the above expansions: In the deep MOND limit, $s\to 0$, we find that
\begin{equation}
Q(r, s) \approx \frac{a_0\sqrt{s}}{r} + \frac{G_0}{2r^2}\frac{dM}{dr}\frac{1}{\sqrt{s}}\, ,
\label{s07}
\end{equation}
and in  the Newtonian limit, $s\to\infty$, we find that
\begin{equation}
Q(r) \approx \frac{2a_0}{r} + \frac{G_0}{r^2}\frac{dM}{dr}\, .
\label{s08}
\end{equation}
These equations delineate the (dis)appearance (and, in some cases, the dominance) of the surface density $s$ in $G(a)$ gravity: $s$ disappears in the deep Newton-Weyl limit and then, the source $Q(r)$ is determined by the gradient $dM/dr$ and the Weyl-like term $2a_0/r$ (eq.~(\ref{s08})). On the other hand, $s$ does not drop out in the deep MOND limit; in fact, the $1/\sqrt{s}$ term becomes dominant in all cases in which the gradient $dM/dr$ takes small or moderate values (eq.~(\ref{s07})).

The function $Q(r, s)$ in eq.~(\ref{s07}) serves as an approximation for the source in Poisson's equation in the deep MOND limit. In fact, the first term is just a small corrrection to the last term since $s\to 0$. When the first term is neglected, and for $dM/dr = 4\pi r^2\rho(r)$, the source takes the compact form
\begin{equation}
Q(r, s) \approx 2\pi G_0\rho(r)/\sqrt{s}\ \ \ \ ({\rm MOND}: ~s\to 0)\, .
\label{s09}
\end{equation}
We see here that the $1/\sqrt{s} > 1$ term amplifies the source in the MOND regime.

On the other hand, the function $Q(r)$ in eq.~(\ref{s08}) serves as the Newton-Weyl approximation for the source in Poisson's equation. The last term, which is precisely equal to $4\pi G_0\rho(r)$, is the familiar Newtonian term. Then eq.~(\ref{s08}) can be written as
\begin{equation}
Q(r) \approx 4\pi G_0 \rho(r) +  \frac{2a_0}{r} \ \ \ \ ({\rm Newton-Weyl}: ~s\to\infty) \, .
\label{s10}
\end{equation}
Once again, we are confronted with the Weyl-like term $2a_0/r$ in the source of the potential that cannot be removed from the deep Newtonian limit except in the absence of modified dynamics ($a_0=0$).

\section{Summary and Discussion}\label{conc}

\subsection{Summary}

In modified gravity with $G(a)$ varying with acceleration $a$, the Newtonian differential form of Gauss's law and Poisson's equation are no longer valid (\S~\ref{df}). The variation of $G$ modifies the source term of these equations irrespective of the choice of the mass distribution $M$ (\S~\ref{Q}). In this work, we have derived this source term in spherical ($r$) symmetry and we have found that the modification is introduced by two functions of the surface density $M(r)/r^2$ that were analyzed in \S~\ref{derivation}. The main result (eq.~(\ref{q2}) or eq.~(\ref{q4})) and the results in special cases in which $M =$ constant or volume density $\rho =$ constant (\S~\ref{cases}), or surface density $M(r)/r^2 =$ constant (\S~\ref{csd})
are not totally unexpected; in the past, there have been many empirical indications that surface density may play an important role at virtually all astrophysical scales \citep{lar81,kaz95,hey09,don09,gen09,lom10,bal12,mil16,tra18}; but this is the first time that $M(r)/r^2$ appears explicitly in the equations and independently of any factors other than $G(a)$ gravity; and the same term disappears conveniently from the equations in the pure Newtonian $G =$ constant case. 

\subsection{Discussion}

The dimensionless surface density $s\propto M(r)/r^2$ persists in the equations of \S~\ref{acc}, even in the simple case of a central point-mass $M =$ constant. In this case, the dimensionless value $s$ naturally reduces to $1/x^2$, where $x$ is a dimensionless distance, and this is probably why its influence was not detected in previous studies. Similarly, the classical Newtonian force law $a_N=G_0M(r)/r^2$ could have been viewed as stating that $a_N\propto s$, thereby indicating the influence of the surface density to classical Newtonian dynamics; unfortunately, this relation has not been previously interpreted in this light, probably because acceleration (a mere second derivative in Newtonian calculus) has never been thought of as a fundamental quantity. Below we discuss the influence of surface density on various terms in our results.

\subsubsection{Interpreting the last term in equation~(\ref{q2})}

Our main result, eq.~(\ref{q2}), reveals an important inference of this study. The last term in the equation for $Q(r)$ can be rewritten as
\begin{equation}
\frac{G_0}{a_0r^2} \frac{dM}{dr}\frac{da}{ds} \ \ \ ({\rm dimension}~[T]^{-2}) \, .
\label{d1}
\end{equation}
This term introduces and couples tightly the acceleration gradient $da/ds$ to the mass gradient $dM/dr$. This coupling is entirely new and it does not occur in Newtonian theory, where the source term $(G_0/r^2)(dM/dr)$ is fully decoupled from $da_N/ds=a_0$. In the deep MOND limit, the new coupling is especially prominent: eq.~(\ref{d1}) (see also eq.~(\ref{s09})) reduces to
\begin{equation}
\frac{G_0}{2r^2} \frac{dM}{dr}\frac{1}{\sqrt{s}}=2\pi G_0\left(\frac{\rho(r)}{\sqrt{s}}\right) \, ,
\label{d2}
\end{equation}
which indicates that the surface density $s\propto M/r^2$ dominates the dynamics in MOND  by amplifying the Newtonian source $\rho(r)$ in the regime $s < 1$ ({\it i.e.}, $\sigma=M/r^2 < a_0/G_0$); in fact, by amplifying $\rho(r)$ acutely as $s\to 0$.

\subsubsection{Interpreting the first term in equation~(\ref{q2})}

The first term on the right-hand side of eq.~(\ref{q2}) also holds a surprise: this term can be rewritten as
\begin{equation}
\frac{2a_0}{r} \frac{d(G/G_0)}{d(1/s)} \ \ \ ({\rm dimension}~[T]^{-2}) \, .
\label{d3}
\end{equation}
Although it was expected that such a term would be created solely by the variation of $G$ with $s$, the surprising element is that the gradient of $G/G_0$ is taken with respect to the reciprocal $1/s$ of the surface density. This realization signifies that $G(s)$ can be viewed as a function of $1/s$ in all regimes. Indeed, eq.~(\ref{f5}) can be recast in the equivalent form
\begin{equation}
G(s) = \frac{G_0}{2}\left(1 + \sqrt{1 + \frac{4}{s}}\right) \, ,
\label{d5}
\end{equation}
which is arguably simpler than eq.~(\ref{f5}) and it is valid for all values of $s$.

\subsubsection{Interpreting the entire source function $Q(r)$}

In all acceleration ($a$) and distance ($r$) regimes, the source function $Q(r)$ given by eq.~(\ref{g5}) or eq.~(\ref{g6}) can be rewritten in the equivalent forms
\begin{equation}
Q(r) = \frac{1}{r^2}\frac{d}{dr}\left( r^2 a \right) = \frac{d}{dV}\left( A a \right) = 4\pi G\rho(r) + \sigma\frac{dG}{dr} \, ,
\label{d6}
\end{equation}
where $\sigma=M(r)/r^2$, $A=4\pi r^2$ is the surface density, and $V=4\pi r^3/3$ is the volume out to distance $r$. The first expression is simply the divergence of vector ${\boldsymbol a}$ in spherical symmetry (see eq.~(\ref{g6})), but it also shows that the gradient of the acceleration is weighted by surface area (the $r^2$ terms). The second expression in eq.~(\ref{d6}) is similar and it shows the precise influence of the curvilinear geometry on to the acceleration $a$. These two expressions are valid in pure Newtonian dynamics as well.
Finally, the last expression in eq.~(\ref{d6}) shows how the the surface density $\sigma$ couples to the gradient $dG/dr$ and it is then introduced into the source function $Q(r)$ by $G(a)$ gravity. Naturally, in the pure Newtonian case ($dG/dr=0$), the same source function assumes the familiar expression $Q(r)=4\pi G_0\rho(r)$ and then $\sigma$ disappears from the source function. Thus, a varying $G$ does not modify the geometric dependencies seen in eq.~(\ref{d6}), it only introduces the surface density via the new term $\sigma(dG/dr)$.

\section*{Acknowledgments}
We are obliged to an anonymous referee for a thorough review of the paper, a review that led to more clarity and some new results pertaining to modified theories of gravity. NASA support over the years is also gratefully acknowledged.

\label{lastpage}

\end{document}